%% file: npa.tex
\documentclass[times,10pt,twocolumn]{article}
\usepackage{url}
\usepackage{graphicx}
\usepackage{tabularx}
\usepackage{fancyhdr}
\usepackage{setspace}
\usepackage{color}
\usepackage{paralist}
\usepackage{wrapfig}
\usepackage{outlines}
\usepackage{latex8}
\usepackage{times}
\usepackage{caption}

\newcommand{\urlBiBTeX}[1]{\url{#1}}
\newcommand{\todo}[1]{}

\begin{document}

\title{Corona: System Implications of Emerging Nanophotonic Technology}
\author{
  Dana Vantrease$^\dag$,
  Robert Schreiber$^\ddag$,
  Matteo Monchiero$^\ddag$,
  Moray McLaren$^\ddag$,
  Norman P. Jouppi$^\ddag$,\\
  Marco Fiorentino$^\ddag$,
  Al Davis$^\S$,
  Nathan Binkert$^\ddag$,
  Raymond G. Beausoleil$^\ddag$,
  Jung Ho Ahn$^\ddag$\\
$^\dag$University of Wisconsin - Madison, $^\ddag$Hewlett-Packard Laboratories, $^\S$University of Utah
}
\maketitle
\thispagestyle{empty}

\begin{abstract}
\input{abstract.tex}
\end{abstract}

\input{intro.tex}
\input{tech.tex}

\input{arch.tex}

\input{method.tex}

\input{eval.tex}

\input{related.tex}
\input{conclusion.tex}

\section*{Acknowledgments}
We thank Ayose Falcon, Paolo Faraboschi, and Daniel Ortega for help with our COTSon simulations.
We also thank Mikko Lipasti and Gabriel Black for their invaluable support and feedback.
Dana Vantrease was supported in part by a National Science Foundation Graduate Research Fellowship.
\singlespacing
\renewcommand{\baselinestretch}{1.00}
{\small
\bibliography{npa}
\bibliographystyle{isca08} 
}

\end{document}

%% file: abstract.tex

We expect that many-core microprocessors will push performance per chip
from the 10 gigaflop to the 10 teraflop range in the coming decade. To
support this increased performance, memory and inter-core bandwidths will
also have to scale by orders of magnitude.  Pin limitations, the energy cost
of electrical signaling, and the non-scalability of chip-length global wires
are significant bandwidth impediments. Recent developments in silicon
nanophotonic technology have the potential to meet these off- and
on-stack bandwidth requirements at acceptable power levels.

Corona is a 3D many-core architecture that uses nanophotonic communication
for both inter-core communication and off-stack communication to memory or
I/O devices. Its peak floating-point performance is 10 teraflops.  Dense
wavelength division multiplexed optically connected memory modules provide
10 terabyte per second memory bandwidth.  A photonic crossbar fully
interconnects its 256 low-power multithreaded cores at 20 terabyte per
second bandwidth.  We have simulated a 1024 thread Corona system running
synthetic benchmarks and scaled versions of the SPLASH-2 benchmark suite.
We believe that in comparison with an electrically-connected many-core
alternative that uses the same on-stack interconnect power, Corona can
provide 2 to 6 times more performance on many memory-intensive workloads,
while simultaneously reducing power.

%% file: intro.tex
\section{Introduction}
\label{sec_intro}

Multi- and many-core architectures have arrived, and core counts are
expected to double every 18 months
\cite{ucb_landscape06}.
As core count grows into the hundreds,
the main memory bandwidth required to support concurrent computation
on all cores will increase by orders of magnitude.
Unfortunately, the ITRS
roadmap~\cite{ITRS} only predicts a small increase in pin count
($<2x$) over the next decade,
and pin data rates are increasing slowly.
This creates a significant bandwidth bottleneck.
Similarly, the inability of
on-chip networks to connect cores to memory or other cores
at the required memory bandwidth poses a serious problem.
Evidence suggests that many-core systems using electrical
interconnects may not be able to meet these high bandwidth demands
while maintaining acceptable performance, power, and
area~\cite{kumar_tullsen}.

Nanophotonics offers an opportunity to reduce the power and area of
off- and on-stack interconnects while meeting future system
bandwidth demands. Optics is ideal for global communication because
the energy cost 
is incurred only at the
endpoints and is largely
independent of length. Dense wavelength division multiplexing (DWDM)
enables multiple single-wavelength communication channels to share a
waveguide, providing a significant increase in bandwidth
density. Recent nanophotonic developments demonstrate that waveguides
and modulation/demodulation circuit dimensions are approaching
electrical buffer and wire circuit dimensions~\cite{lipson05}.

Several benefits accrue when nanophotonics is coupled to emerging 3D
packaging~\cite{3D_wkshop}.  The 3D approach allows multiple die, each
fabricated using a process well-suited to it, to be stacked and to
communicate with through silicon vias (TSVs). Optics, logic, DRAM,
non-volatile memory (e.g. FLASH), and analog circuitry may all occupy
separate die and co-exist in the same 3D package.  Utilizing the third
dimension eases layout and helps decrease worst case wire lengths.

Corona is a nanophotonically connected 3D
many-core NUMA system that meets the future bandwidth demands of
data-intensive
applications at acceptable power levels. Corona
is targeted for a 16 nm process in 2017.  Corona comprises
256 general purpose cores, organized in 64 four-core clusters, and is
interconnected by an all-optical, high-bandwidth DWDM crossbar. The
crossbar enables a cache coherent design
with near uniform on-stack and memory communication latencies.
Photonic connections to off-stack memory enables unprecedented
bandwidth to large amounts of memory with only modest power
requirements.

This paper presents key aspects of nanophotonic technology, and
considers the implications for many-core
processors. It describes the Corona architecture, and presents a
performance comparison to a comparable, all-electrical many-core
alternative.  The
contribution of this work is to show that
nanophotonics is compatible with future CMOS technology,
is capable of dramatically better communication performance
per unit area and energy,
and can significantly improve the performance and utility of
future many-core architectures.

%% file: tech.tex
\section{Photonic Technology}
\label{sec_technology}

Advances in silicon nanophotonics have made complete photonic
on-stack communication networks a serious alternative to electrical
networks. Photonic interconnects are interesting because they can be
much more energy efficient than their electrical counterparts,
especially at high speeds and long distances. In addition, the
ability of optical fibers and waveguides to carry many information
channels simultaneously increases interconnect bandwidth density
significantly and eliminates the need for a large number
of wires in order to achieve adequate bandwidth.  Photonics dominates long-haul interconnects and is
increasingly ubiquitous in metropolitan, storage, and local area networks. Photonic interconnects are
becoming standard in data centers, and  chip-to-chip optical links have been demonstrated \cite{terabus}.   This trend will naturally bring photonic interconnects
into the chip stack, particularly since limited pin density and the power
dissipation of global wires places
significant constraints on performance and power.

A complete nanophotonic network requires waveguides to carry signals, light
sources that provide the optical carrier, modulators that encode the data
onto the carrier, photodiodes to detect the data, and injection switches
that route signals through the network. It is imperative that the optical
components be built in a single CMOS-compatible process to reduce the cost of
introducing this new technology.  Waveguides confine and guide light and
need two optical materials: a high refraction index material to
form the core of the waveguide and a low index material that forms the
cladding. We use crystalline silicon (index $\approx 3.5$) and silicon oxide (index
$\approx 1.45$).  Both are commonly used in CMOS processes.  A silicon oxide
waveguide has typical cross-sectional dimensions of $\approx500$ nm with a
wall thickness of least 1 $\mu$m. These waveguides have been shown to be
able to carry light with losses on the order of 2--3~dB/cm and can be
curved with bend radii on the order of 10~$\mu$m.

In order to communicate rather than simply illuminate, we must introduce a
second material to absorb light and convert the light into an electric
signal. Germanium is a natural choice: it is already being used in CMOS
processes and has a significant photo-absorption between 1.1 and 1.5
$\mu$m.  While it is possible to induce strains in the crystalline
structure of Ge to extend its absorption window into the 1.55 $\mu$m window
commonly used for long distance fiber communication, it is easier to rely
on unstrained Ge and use light around 1.3 $\mu$m.  This wavelength is still
compatible with fiber transmission, which is an important characteristic
for off-stack networks that will need to use some kind of waveguide or
optical fiber to connect different optically enabled stacks.

The third element is a light source, and a laser is the obvious choice.  A
laser's narrow linewidth allows one to pack many communication channels in
a single waveguide, thus increasing bandwidth.  There are two possible ways
to encode data in laser light.  The first method uses {\em direct} modulation of
the laser, where the laser is switched on and off to represent digital 1s
and 0s.  The second method uses a continuous-wave (CW) laser and a {\em separate}
modulator to achieve the required modulation.  To achieve the high
modulation speeds that would make on-stack interconnects practical
(typically 10 Gb/s data rates) one would need to use vertical cavity
semiconductor lasers (VCSELs) for direct modulation.  Since VCSELs are
built using III-V compound semiconductors, they cannot be easily integrated
in a CMOS-compatible process.  On-stack mode-locked lasers are an
interesting separate modulation alternative.  A mode-locked laser generates
a comb of phase-coherent wavelengths at equally spaced wavelengths.
On-stack mode-locked lasers have been recently demonstrated~\cite{bowers}.
It is expected that one such laser could provide 64 wavelengths for a
DWDM network.

\begin{figure*}[t!]
\begin{center}
\includegraphics[width=6.0in]{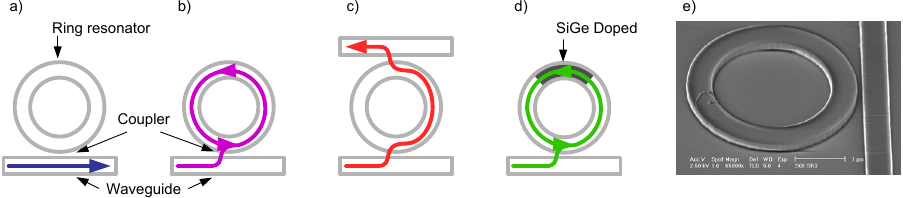}
\caption{
    Schematic Top View of Nanophotonic Building Blocks.
       \small{
   {\bf(a)  Off-resonance Modulator.} The ring resonator is coupled to a waveguide through evanescent coupling.
   Off-resonance wavelengths are transmitted through.
   {\bf(b)  On-resonance Modulator.}  A resonant wavelength is coupled in the ring and eventually gets attenuated by losses.
   A negligible amount of light is transmitted through due to destructive interference.  By switching between the on- and off-resonance state one can achieve modulation (or diversion)
   of a continuous-wave laser.
   {\bf(c)  Injector.}  A resonant wavelength in the input (lower) waveguide is coupled into the ring and out through the output
   (top) waveguide.
   {\bf(d)  Detector.}  A resonant wavelength is coupled in the ring and absorbed by a SiGe detector coupled to the ring.
   {\bf(e) } SEM image of a 3 $\mu$m diameter resonator ring. Image courtesy of Qianfan Xu,  HP Labs Palo Alto.
   \label{fig_f:rings}}}
\end{center}
\end{figure*}

For separate modulation, external modulators
are required.  In a DWDM network it is preferable to have
wavelength-selective modulators that can modulate a
single wavelength in a multi-wavelength channel.  This simplifies the topology of the network and increases its
flexibility. Wavelength-selective silicon modulators with
modulation rates in excess of 10 Gb/s have recently been
demonstrated~\cite{qianfan}.  These modulators are based on ring
resonators built using an SOI waveguide in a ring with
diameter of 3-5 $\mu$m~(Figure~\ref{fig_f:rings}(a)).
Depending on their construction, they may serve to
modulate, inject, or detect the light.

To modulate light, a ring is placed next to a waveguide.  A fraction of the
light will be coupled and circulate inside the ring. For some wavelengths,
the length of the ring circumference will be equal to an integer number of
wavelengths. In this resonance condition, the light in the ring will be
enhanced by interference while the light transmitted by the waveguide will
be suppressed~(Figure~\ref{fig_f:rings}(b)).  In a properly coupled ring, the
waveguide's light will be suppressed and the rings' light will be
eventually lost due
to the bend in the ring and scattering from imperfections.  The wavelength
depends primarily on the index of refraction of the materials that
constitute the ring.  We can further fine-tune the wavelength (or index of
refraction) by injecting charge into the ring or changing the temperature
of the ring.  This brings the ring in and out of resonance.  The first
method is commonly used for fast modulation while the second can be used
for slow tuning.

The same ring structure can be used to inject a single wavelength from one
waveguide to another.  If the ring is placed between two waveguides and the
coupling between the ring and the waveguides is properly chosen, the
wavelength resonating within the ring will be transferred from one
waveguide to the other.  By bringing the ring in and out of resonance,
a frequency-selective switch injector ~(Figure~\ref{fig_f:rings}(c)) can be
realized.

A ring resonator can also be used as a wavelength-selective
detector~(Figure~\ref{fig_f:rings}(d)).  If germanium is included in the
resonator, the resonant wavelength will be absorbed by the germanium and it
will generate a photocurrent while non-resonant wavelengths will be
transmitted.  An advantage of this scheme is that because the resonant
wavelength will circulate many times in the ring,
only very small absorption rates (less than 1\% per pass) will be needed and
therefore a small detector will be sufficient.  This brings
the capacitance of the detector down to $\approx$ 1 fF and removes the need
for power-hungry trans-impedance amplifiers.


A final component that is not necessary for photonic networks but that we
find useful is a broadband splitter.  A broadband splitter distributes
power and data by splitting the signal between two waveguides.  It diverts
a fixed fraction of optical power from all wavelengths of one waveguide and
injects them onto another waveguide.  Other than a drop in strength, the
unsplit portion of the signal is unaffected by the splitter.

While most of the individual components of a DWDM on-stack network have
been demonstrated~\cite{bowers,qianfan}, a number of important problems
remain. Foremost among these is the necessity to integrate a large number
of devices in a single chip.  It will be necessary to analyze and correct
for the inevitable fabrication variations to minimize device failures and
maximize yield.  A large effort will also be needed to design the analog
electronics that drive and control the optical devices in a power-efficient
way. While significant research is still necessary, we believe that DWDM
photonic networks offer a credible answer to the challenges posed by the
increasing bandwidth needs of many-core architectures.

%% file: arch.tex
\section{The Corona Architecture}
\label{sec_arch}

\begin{figure}[t!]
\center
\includegraphics[width=.33\textwidth]{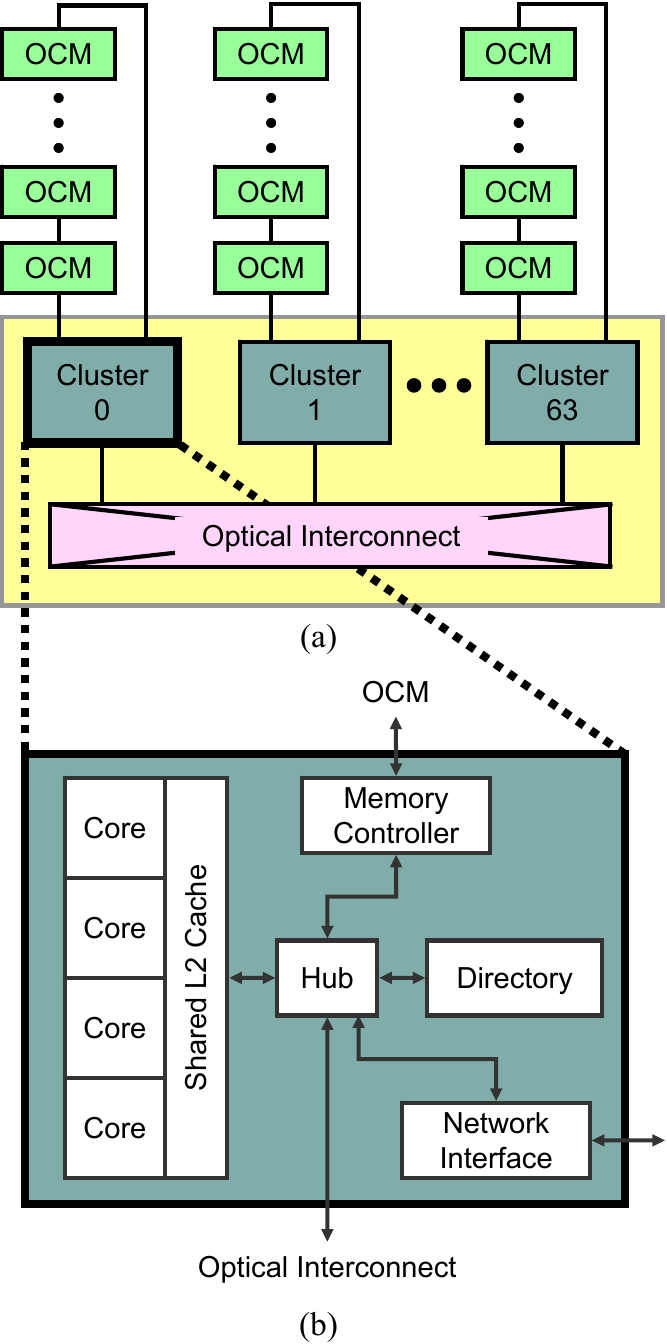}
\caption{Architecture Overview}
\label{fig_archsum}
\end{figure}

Corona is a tightly coupled, highly parallel NUMA system. As NUMA systems
and applications scale, it becomes more difficult for the
programmer, compiler, and runtime system to manage the placement and migration
of programs and data.  We try to lessen the burden with
homogeneous cores and caches, a crossbar interconnect that has
near-uniform latency, a fair interconnect arbitration protocol, and high
(one byte per flop) bandwidth between cores and from caches to
memory.

The architecture is made up of 256 multithreaded in-order cores and
is capable of supporting up to 1024 threads simultaneously,
providing up to 10 teraflops of computation,
up to 20 terabytes per second (TB/s) of on-stack bandwidth,
and up to 10 TB/s of off-stack memory bandwidth.

Figure~\ref{fig_archsum} gives a conceptual view of the system while
Figure~\ref{fig_explode} provides a sample layout of the system including the waveguides that comprise the optical interconnect (Section~\ref{sec_onstack_interconnect}), the optical connection to memory (Section~\ref{sec_optmem}), and other optical components.

\subsection{Cluster Architecture}
\label{sec_cores}

\begin{figure*}[ht!]
\center
\includegraphics[width=.79\textwidth]{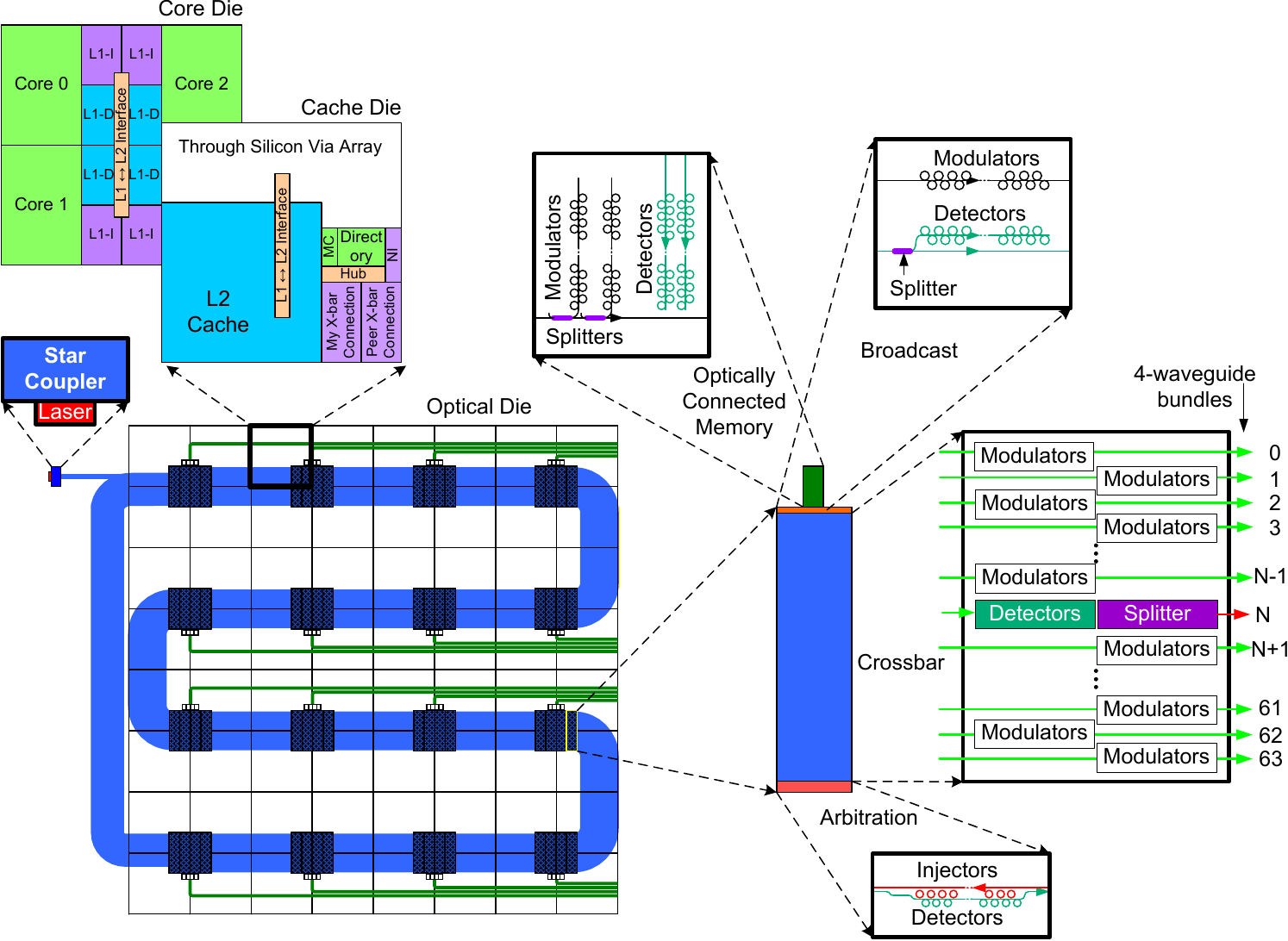}
\caption{Layout with Serpentine Crossbar and Resonator Ring Detail}
\label{fig_explode}
\end{figure*}

Each core has a private L1 instruction and data cache, and
all four cores share a unified L2 cache. A hub routes message
traffic between the L2, directory, memory controller, network
interface, optical bus, and optical crossbar.
Figure~\ref{fig_archsum}(b) shows the cluster configuration, while
the upper left hand insert in Figure~\ref{fig_explode}
shows its approximate floorplan.

Because Corona is an architecture targeting future high throughput systems,
our exploration and evaluation of the architecture
is {\it not} targeted at
optimal configuration of the clusters' subcomponents (such as their branch prediction
schemes, number of execution units, cache sizes, and cache policies).
Rather, the clusters' design parameters (Table \ref{tab_cpr})
represent reasonably modest choices for a
high-performance system targeted at a 16 nm process in 2017. 

\begin{table}[t!]
\small
\center{
\begin{tabular}{l|r}
{Resource}  & {Value} \\
\hline
Number of clusters & 64 \\
Per-Cluster: & \\
\hspace{.15in} L2 cache size/assoc & $4$ MB/$16$-way\\
\hspace{.15in} L2 cache line size & $64$ B\\
\hspace{.15in} L2 coherence & MOESI\\
\hspace{.15in} Memory controllers & $1$\\
\hspace{.15in} Cores & $4$  \\
\hspace{.15in} Per-Core: & \\
\hspace{.3in} L1 ICache size/assoc & $16$ KB/$4$-way\\
\hspace{.3in} L1 DCache size/assoc & $32$ KB/$4$-way\\
\hspace{.3in} L1 I \& D cache line size & $64$ B\\
\hspace{.3in} Frequency & $5$ GHz \\
\hspace{.3in} Threads & $4$ \\
\hspace{.3in} Issue policy & In-order \\
\hspace{.3in} Issue width & $2$ \\
\hspace{.3in} 64 b floating point SIMD width & $4$ \\
\hspace{.3in} Fused floating point operations & Multiply-Add \\
\end{tabular}
}
\caption{Resource Configuration}
\label{tab_cpr}
\end{table}

\subsubsection{Cores}
The core choice is primarily motivated by power; with
hundreds of cores, each one will need to be extremely energy
efficient.
We use dual-issue, in-order, four-way multithreaded cores.

Power analysis has been based on the Penryn~\cite{penryn} (desktop and
laptop market segments) and the Silverthorne ~\cite{silverthorne}
(low-power embedded segment) cores.
Penryn is a single-threaded out-of-order core
supporting 128-bit SSE4 instructions.  Power per core has been conservatively
reduced by $5x$ (compared to the $6x$ predictions in \cite{ucb_landscape06}) and then
 increased by 20\% 
to account for differences in the quad-threaded Corona.
Silverthorne is a dual-threaded in-order 64-bit design where power and area
have been increased to account for the Corona architectural
parameters.
Directory and L2 cache power has been calculated using CACTI 5~\cite{cacti5}.
Hub and memory controller power estimates are based on synthesized 65 nm
designs and Synopsis Nanosim power values scaled to 16 nm.
Total processor, cache, memory controller and
hub power for the Corona design is expected to be between
82 watts (Silverthorne based) and 155 watts (Penryn based).


Area estimates are based on pessimistically scaled Penryn and
Silverthorne designs.  We estimate that an in-order Penryn core will have
one-third the area of the existing out-of-order Penryn.  This estimate is
consistent with the current core-size differences in 45 nm for the
out-of-order Penryn and the in-order Silverthorne, and is
more conservative than the $5x$ area reduction reported by Asanovic et
al. \cite{ucb_landscape06}.  We then assume a multithreading area overhead
of 10\% as reported in Chaudry et al. \cite{chaudhry05}.  Total die area
for the processor and L1 die is estimated to be between 423 mm$^{2}$
(Penryn based) and 491 mm$^{2}$ (Silverthorne based).  The discrepancy
between these estimates is likely affected by the 6-transistor Penryn L1
cache cell design vs. the 8-transistor Silverthorne L1 cache cell.

\subsubsection{On-Stack Memory}
Corona employs a MOESI directory protocol.
The protocol is backed by a single broadcast bus (see
Section~\ref{sec_opt_bcast}), which
is used to quickly invalidate a large pool of sharers with a single message.
The coherence
scheme was included for purposes of die size and power estimation,
but has not yet been modeled in the system simulation.
Nonetheless, we believe that our
performance simulations provide
a sensible first-order indication of Corona's potential.

The Corona architecture has one memory controller per cluster.
Associating the memory controller with the cluster ensures that the
memory bandwidth grows linearly with increased core count, and it provides local memory accessible with low latency.
Photonics connects the memory controller to off-stack memory as
detailed in Section~\ref{sec_optmem}.

Network interfaces, similar to the interface to off-stack main
memory, provide inter-stack communication for larger systems using
DWDM interconnects.

\subsection{On-Stack Photonic Interconnect}
\label{sec_onstack_interconnect}

Corona's 64 clusters communicate through an optical
crossbar and occasionally an optical broadcast ring.
Both are managed using optical tokens.
Several messages of
different sizes may simultaneously share any communication channel,
allowing for high utilization.

Table~\ref{tab_xbar}
summarizes the interconnects' optical component requirements
(power waveguides and I/O components are omitted).
Based on existing designs, we
estimate that the photonic interconnect power (including the power
dissipated in the analog circuit layer and the laser power in the
photonic die) to be 39 W.


\begin{table}[tb]
\small
\center{
    \begin{tabular}{l|r|r}
    { Photonic Subsystem} & { Waveguides}  & { Ring Resonators} \\
    \hline
    Memory      & $128$ & $16$ K\\
    Crossbar    & $256$ & $1024$ K\\
    Broadcast   & $1$   & $8$ K\\
    Arbitration & $2$   & $8$ K\\
    Clock       & $1$   & $64$\quad\quad\\
    {\bf$\overline{\mathrm{Total}\quad\quad}$}      & {\bf$\overline{388}$} & $\overline{\approx1056\:\mathrm{K}}$ \\
     \end{tabular}
}
\caption{Optical Resource Inventory}
\label{tab_xbar}
\end{table}

\subsubsection{Optical Crossbar}
\label{sec_onstack_crossbar}
Each cluster has a designated channel that address, data, and
coherence messages share.
Any cluster may write to a given channel, but only a single fixed cluster
may read from the channel.
A fully-connected $64\times64$ crossbar can be realized by replicating this
many-writer single-reader channel 64 times, adjusting the assigned
``reader'' cluster with each replication.

The channels are each 256 wavelengths, or 4 bundled waveguides, wide.
When laid out, the waveguide bundle forms a broken ring that originates at the
destination cluster (the channel's {\it home} cluster), is routed
past every other cluster, and eventually terminates back at its
origin. Light is sourced at a channel's home by a splitter
that provides all wavelengths of light from a power
waveguide. Communication is unidirectional, in cyclically increasing
order of cluster number.

\begin{figure}[t!]
\center
\includegraphics[width=.45\textwidth]{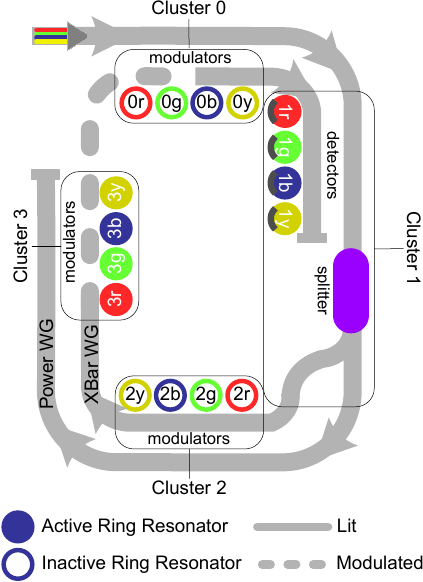}
\caption{
    A Four Wavelength Data Channel Example.
        \small{
    The home cluster (cluster 1) sources all wavelengths of light (r,g,b,y).
    The light travels clockwise around the crossbar waveguides.
    It passes untouched by cluster 2's inactive (off-resonance) modulators.
    As it passes by cluster 3's active modulators, all wavelengths
    are modulated to encode data.
    Eventually, cluster 1's detectors sense the modulation, at which
    point the waveguide terminate.
        }
    }
\label{fig_xbar}
\end{figure}

A cluster sends to another cluster by modulating the light on the
destination cluster's channel.  Figure~\ref{fig_xbar} 
illustrates the conceptual operation of a four-wavelength channel.  Modulation occurs
on both clock edges, so that each of the wavelengths signals at 10 Gb/s,
yielding a per-cluster bandwidth of 2.56 terabits per second (Tb/s) and a
total crossbar bandwidth of 20 TB/s.

A wide phit with low modulation time
keeps latency to a minimum, which is critical to ensuring the
in-order cores minimize stall time. A 64-byte cache line can be
sent (256 bits in parallel twice per clock) in one 5 GHz clock.
The propagation time is at most 8 clocks and is determined by a combination
of the source's distance from the destination and the speed of
light in a silicon waveguide (approximately 2 cm per clock).
Because messages, such as cache lines, are localized
to a small portion of
the bundle's length, a bundle may have multiple back-to-back
messages in transit simultaneously.



Corona uses optical global distribution of the clock in order to
avoid the need for signal retiming at the destination.  A clock
distribution waveguide parallels the data waveguide, with the clock
signal traveling clockwise with the data signals. This means that
each cluster is offset from the previous cluster by approximately
1/8th of a clock cycle. A cluster's electrical clock is phase
locked to the arriving optical clock. Thus, input and output data
are in phase with the local clock; this avoids costly retiming
except when the serpentine wraps around.


\subsubsection{Optical Broadcast Bus}
\label{sec_opt_bcast}
In the MOESI coherency protocol, when a shared block is invalidated,
an invalidate message must be multicast to all sharers.
For unicast interconnects, such as Corona's crossbar, the multicast is translated into several unicast messages.
These unicast messages cause network congestion and may harm performance~\cite{virtual_multicast}.

We avoid redundant unicast invalidates by augmenting our system with a broadcast bus.
The broadcast bus is a single waveguide that passes by each cluster twice in a coiled, or spiral-like, fashion.
The upper right hand corner of Figure~\ref{fig_explode}
gives a cluster-centric view of the bus' details.
The light is sourced from one endpoint of the coil.
On the light's first pass around the coil, clusters modulate the light to encode invalidate messages.
On the light's second pass, the invalidate messages become active, that is, clusters may read the messages and snoop their caches.
To do this, each cluster has a splitter that transfers a fraction of the light from
the waveguide to a short dead-end waveguide that is populated with detectors.

In addition to broadcasting invalidates, the bus' functionality could be generalized for other broadcast applications, such as bandwidth adaptive snooping \cite{adaptive_snooping} and barrier notification.

\subsubsection{Arbitration}
\label{sec_arb}

\begin{figure}[ht!b]
    \center
    \includegraphics[width=.45\textwidth]{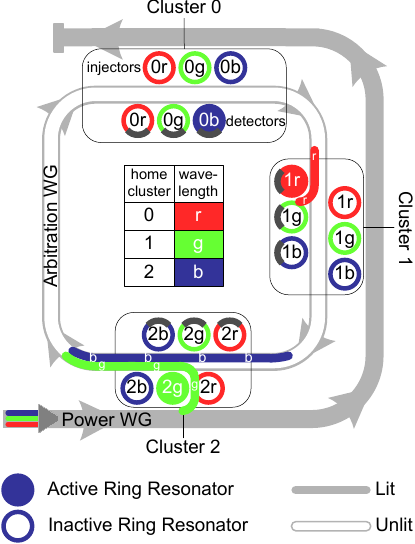}
    \caption{
        Optical Token Arbitration Example.
            \small{
        3 wavelengths are used to arbitrate for 3 channels.
        The channel-to-token (home cluster-to-wavelength) mapping is shown
        in the embedded table.
        In this depiction, all tokens are in transit (i.e. all channels are free).
        {\bf Cluster 0} is requesting cluster 2 (blue), will soon divert
         blue, and will then begin transmitting on cluster 2's channel.
        {\bf Cluster 1} is requesting cluster 0 (red), has nearly completed diverting red, and will soon begin transmitting on cluster 0's channel.
        {\bf Cluster 2} is re-injecting green (cluster 1) because it has just finished
transmitting on cluster 1's channel.
        (Note: Detectors are positioned to prevent a cluster from re-acquiring
        a self-injected token until it has completed one revolution around the ring.)
            }
    }
    \label{fig_arbitration}
\end{figure}

The crossbar and broadcast bus both require a conflict resolution scheme
to prevent two or more sources from concurrently sending to the same destination.
Our solution is a distributed, all optical, token-based
arbitration scheme that fairly allocates the available interconnect
channels to clusters. Token ring arbitration is naturally
distributed and has been used in token ring LAN systems
\cite{802p5}. Its asynchronous acquire-and-release nature tolerates
variability in request arrival time,
message modulation time, and message propagation time.
Figure~\ref{fig_arbitration} demonstrates a photonic version of this approach.

In our implementation, a token conveys the right to send data on a given
channel.
The one-bit token is
represented by the presence of a short signal in a specific
wavelength on an arbitration waveguide. For the broadcast bus,
we use one wavelength.
For the crossbar, we use 64 wavelengths, in
one to one correspondence with the 64 crossbar channels.

When a cluster finishes sending a message on a channel, it releases
the channel by activating its injector and re-introducing the token
onto the arbitration waveguide.
The token travels in parallel with the tail of the most recently transmitted message.
Each cluster is
equipped with an array of fixed-wavelength detectors that are
capable of diverting (obtaining) any token. If a token is diverted,
the light is completely removed from the arbitration waveguide to
provide an exclusive grant for the corresponding channel.
Each cluster
will absorb and regenerate its channel token to
ensure that it remains optically sound even after many trips around
the ring without any ``takers.''

This scheme fairly allocates the channels in a
round-robin order.
When many clusters want the same channel and contention is high,
token transfer time is low and channel utilization is high.
However when contention is low, a cluster may wait as long as
8 processor clock cycles for an uncontested token.

\subsection{Optically Connected Memory}
\label{sec_optmem}
One design goal is to scale main memory
bandwidth to match the growth in computational power. Maintaining
this balance ensures that the performance of the system is not
overly dependent on the cache utilization of the application. Our
target external memory bandwidth for a 10 teraflop processor is 10 TB/s.
Using an electrical interconnect to achieve this performance would require
excessive power; over 160 W assuming 2 mW/Gb/s~\cite{rambus2007tranceiver}
interconnect power. Instead, we use a nanophotonic interconnect that has high
bandwidth and low power. The same
channel separations and data rates that are used on the internal
interconnect network can also be used for external fiber connections. We
estimate the interconnect power to be 0.078 mW/Gb/s, which equates
to a total memory system power of approximately 6.4 W.

Each of the 64 memory controllers connects to its external memory by
a pair of single-waveguide, 64-wavelength DWDM links. The optical
network is modulated on both edges of the clock. Hence each
memory controller provides 160 GB/s of off-stack memory bandwidth, and
all memory controllers together provide 10 TB/s.

This allows all communication to be
scheduled by the memory controller with no arbitration.
Each external optical communication link consists of a
pair of fibers providing half duplex communication between the CPU
and a string of optically connected memory (OCM) modules. The
link is optically powered from the chip stack; after connecting to the
OCMs, each outward fiber is looped back as a return fiber.
Although the off-stack memory interconnect uses the same modulators and
detectors as the on-stack interconnects, the communication protocols differ.
Communication between processor and memory is master/slave, as
opposed to peer-to-peer.
To transmit, the memory controller modulates the light and the target module
diverts a portion of the light to its detectors.
To receive, the memory controller detects light that the transmitting
OCM has modulated on the return fiber.  Because the memory controller is the
master, it can supply the necessary unmodulated power to the transmitting OCM.

\begin{figure*}[t!]
\center
$\begin{array}{c@{\hspace{.10in}}c@{\hspace{.10in}}c@{\hspace{.02in}}c}
 \includegraphics[width=2.4in]{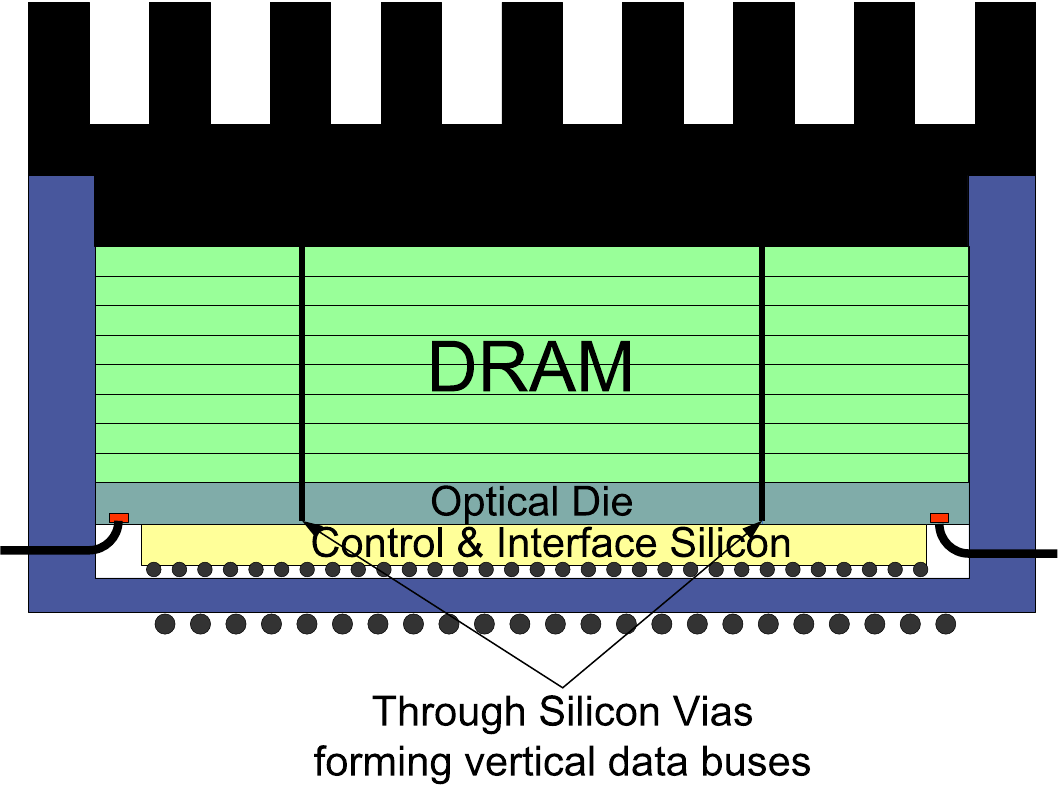}  &
 \includegraphics[width=1.6in]{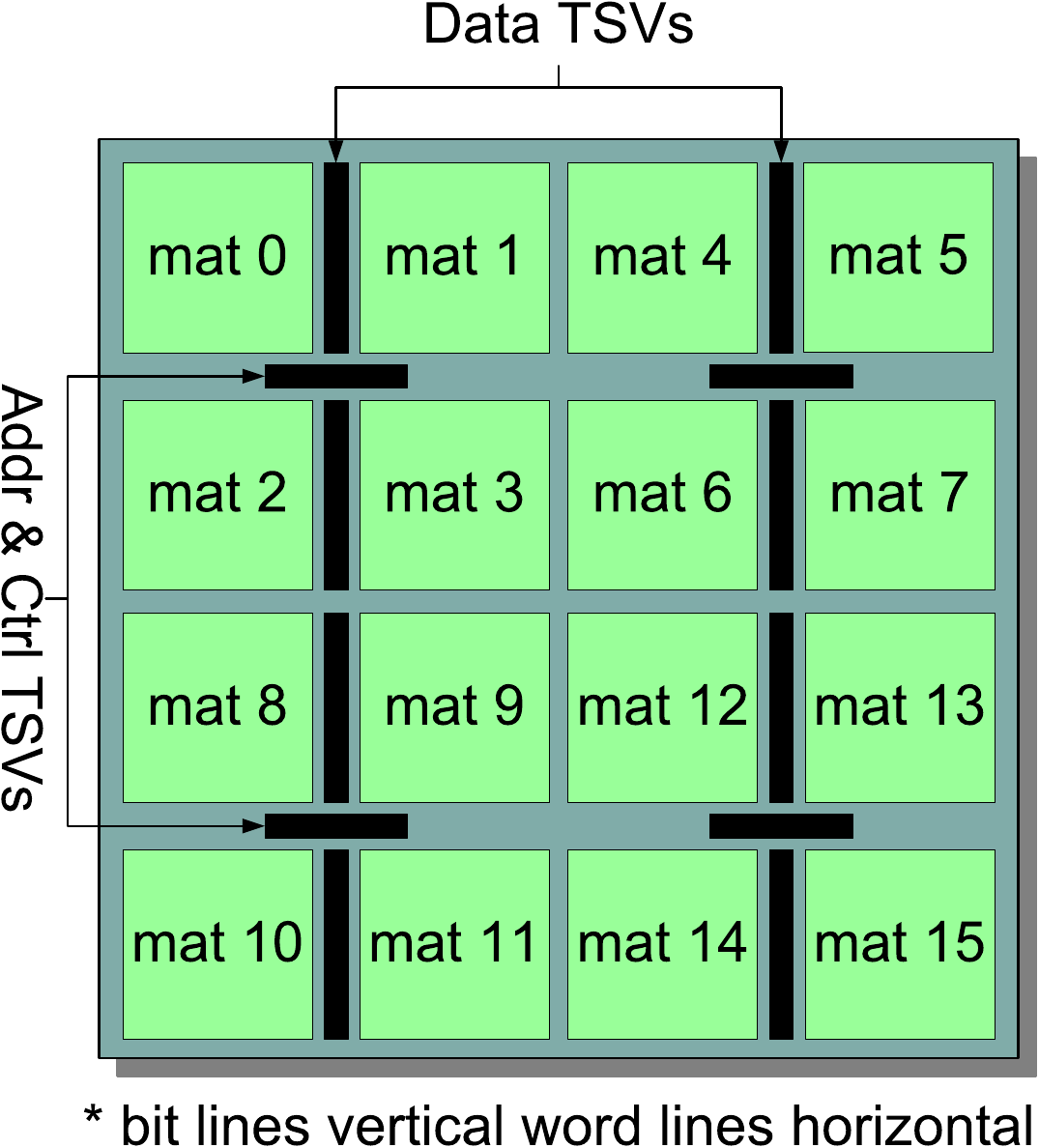}  &
 \includegraphics[width=2.3in]{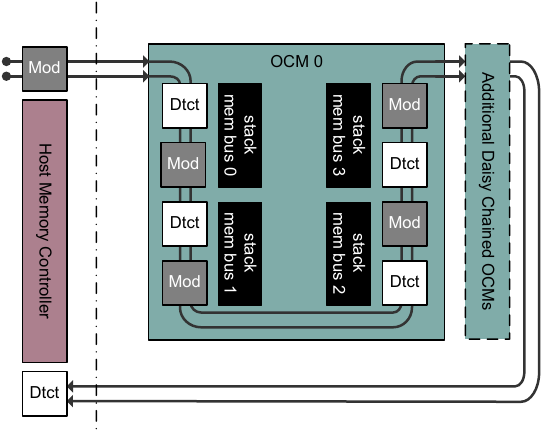}  &  \\
                        $(a)$ & $(b)$ & $(c)$\\
                       \end{array}$
\caption{
Schematic View of Optically Connected Memory.
\small {{\bf(a) 3D die stack.} The stack has one optical die and
multiple DRAM dies.
{\bf(b) DRAM die floorplan.}  Each quadrant is
independent, and could also be constructed from four independent
die.
{\bf (c) OCM expansion.}  The light travels from the processor,
through one or more OCMs, finally looping back to the processor.}}
\label{fig_ocm}
\end{figure*}

Figure~\ref{fig_ocm}(a) shows the 3D stacked OCM module, built from
custom DRAM die and an optical die. The DRAM die is organized so that
an entire cache line is read or written from a single mat.
3D stacking is used to
minimize the delay and power in the interconnect between the optical
fiber loop and the DRAM mats. The high-performance
optical interconnect allows a single mat to quickly provide all the
data for an entire cache line. In contrast, current electrical memory systems
and DRAMs activate many banks on many die on a DIMM, reading out
tens of thousands of bits into an open page. However, with highly
interleaved memory systems and a thousand threads, the chances of
the next access being to an open page are small. Corona's DRAM architecture
avoids accessing an order of
magnitude more bits than are needed for the cache line, and hence
consumes less power in its memory system.

Corona supports memory expansion by adding additional OCMs
to the fiber loop as shown in Figure~\ref{fig_ocm}(c). Expansion
adds only modulators and detectors and not lasers, so the
incremental communication power is small.  As the light passes
directly through the OCM without buffering or retiming, the
incremental delay is also small, so that the memory access latency is
similar across all modules.  In contrast, a serial electrical scheme,
such as FBDIMM, would typically require the data to be resampled and
retransmitted at each module, increasing the communication power and
access latency.

\subsection{Chip Stack}

\begin{figure}[ht!]
\center
\includegraphics[width=.45\textwidth]{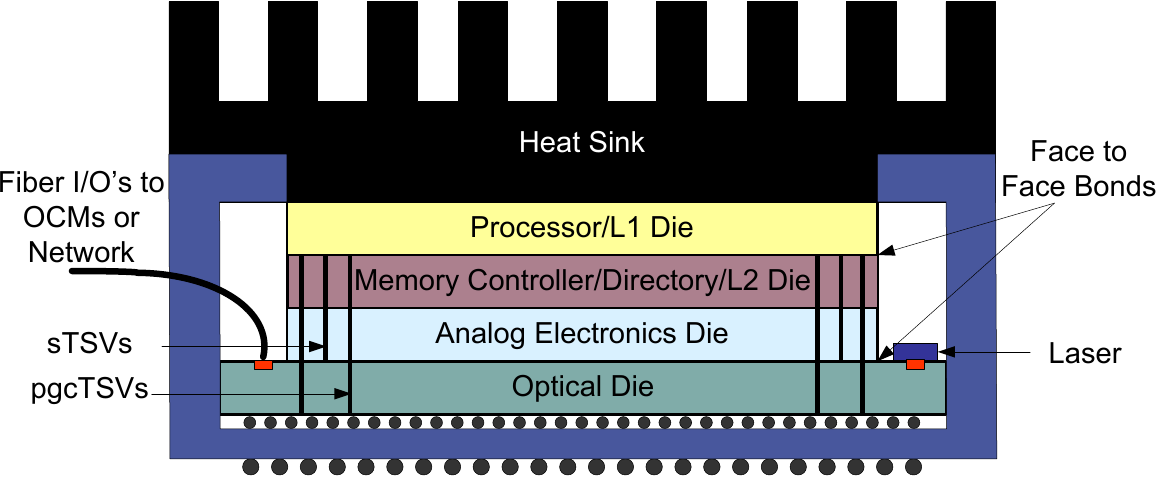}
\caption{Schematic Side View of 3D Package}
\label{fig_npa_stack}
\end{figure}

Figure~\ref{fig_npa_stack} illustrates the Corona 3D die stack. Most of the
signal activity, and therefore heat, are in the top die (adjacent to the
heat sink) which contains the clustered cores and L1 caches.  The processor
die is face-to-face bonded with the L2 die, providing direct connection
between each cluster and its L2 cache, hub, memory controller, and
directory.  The bottom die contains all of the optical structures
(waveguides, ring resonators, detectors, etc.) and is face-to-face bonded
with the analog electronics which contain detector circuits and control
ring resonance and modulation.

All of the L2 die components are
potential optical communication end points and connect to the analog die by
signal through silicon vias (sTSVs).  This strategy minimizes the layout
impact since most die-to-die signals are carried in the face-to-face bonds.
External power, ground, and clock vias (pgcTSVs) are the only TSVs that
must go through three die to connect the package to the top two digital
layers.  The optical die is larger than the other die in order to expose a
{\it mezzanine} to permit fiber attachments for I/O and OCM channels and
external lasers.

%% file: method.tex
\section{Experimental Setup}
\label{sec_method}
We subject our architecture to a combination of synthetic and realistic workloads that were selected with an eye to stressing the on-stack and memory interconnects.
Synthetic workloads stress particular features and aspects of the interconnects.
The SPLASH-2 benchmark suite~\cite{splash2} indicates their realistic performance.
The SPLASH-2 applications are not modified in their essentials.
We use larger datasets
when possible to ensure that each core has a nontrivial workload.
Because of a limitation in our simulator, we needed to replace implicit
synchronization via semaphore variables with explicit synchronization constructs.
In addition, we set the L2 cache size to 256 KB to better match our simulated
benchmark size and duration when scaled to expected system workloads.
A summary of the workload setup is described in Table \ref{tab_bmarks}.

\begin{table}
\small
\center{
\begin{tabular}{l|r@{ }l|r}
\multicolumn{4}{l}{\bf Synthetic}  \\
{         }   & \multicolumn{2}{l|}{           } & {\# Network} \\
{Benchmark}   & \multicolumn{2}{l|}{Description} & {Requests}  \\
        \hline
Uniform   & \multicolumn{2}{l|}{Uniform random }              & 1 M \\
Hot Spot  & \multicolumn{2}{l|}{All clusters to one cluster}  & 1 M \\
Tornado   & \multicolumn{2}{l|}{Cluster $(i,j)$ to cluster}   & 1 M \\
          & \multicolumn{2}{l|}{\quad $((i+\lfloor k/2 \rfloor -1)\%k,$} & \\
          & \multicolumn{2}{l|}{\quad $\ (j+\lfloor k/2 \rfloor -1)\%k),$} & \\
          & \multicolumn{2}{l|}{\quad where $k=$ network's radix} & \\
Transpose & \multicolumn{2}{l|}{Cluster $(i,j)$ to cluster $(j,i)$}& 1 M \\
\multicolumn{4}{l}{}\\
\multicolumn{4}{l}{\bf SPLASH-2 }  \\
{         }  & \multicolumn{2}{c|}{Data Set}               & {\# Network} \\
{Benchmark}  & \multicolumn{2}{c|}{Experimental (Default)} & {Requests}  \\
        \hline
 Barnes    & 64 K particles & (16 K)                         & 7.2 M\\
 Cholesky  & tk29.O        & (tk15.O)                      & 0.6 M\\
 FFT       & 16 M  points   & (64 K)                         & 176 M\\
 FMM       & 1 M particles  & (16 K)                         & 1.8 M\\
 LU        & 2048$\times$2048 matrix & (512$\times$512)    &  34 M\\
 Ocean     & 2050$\times$2050 grid   & (258$\times$258)    & 240 M\\
 Radiosity & roomlarge     & (room)                        & 4.2 M\\
 Radix     & 64 M  integers & (1 M)                          & 189 M\\
 Raytrace  & balls4        & (car)                         & 0.7 M\\
 Volrend   & head          & (head)                        & 3.6 M\\
 Water-Sp  & 32 K molecules & (512)                         & 3.2 M\\
\end{tabular}
}
\caption{Benchmarks and Configurations}
\label{tab_bmarks}
\end{table}

The simulation infrastructure is split into two independent
parts: a full system simulator for generating L2 miss memory
traces and a network simulator for processing these traces.
A modified version of the HP Labs' COTSon simulator~\cite{falconISPASS07}
generates the traces.
(COTSon is based on AMD's SimNow simulator infrastructure.)
Each application is compiled with gcc 4.1, using -O3 optimization, and run as a single
1024-threaded instance.
We are able to collect multithreaded traces by translating the operating system's thread-level parallelism into hardware thread-level parallelism.
In order to keep the trace files and network simulations manageable,
the simulators do not tackle the intricacies of cache coherency between clusters.

The network simulator reads the traces and processes them in the network subsystem.
The traces consist of L2 misses and synchronization events that are annotated with thread id and timing information.
In the network simulator, L2 misses go through a request-response,
on-stack interconnect transaction and an off-stack memory transaction.
The simulator, which is based on the M5 framework~\cite{m5}, takes an
trace-driven approach to processing memory requests.
The MSHRs, hub, interconnect, arbitration, and memory
are all modeled in detail with finite buffers, queues, and ports.
This enforces bandwidth, latency, back pressure, and capacity limits throughout.
\newcommand{\xbarlatency}{8} 
\newcommand{\xbarbbw}{20.48 TB/s}
\newcommand{\hmeshbbw}{{\bf {\it TBD \ }}}
\newcommand{\lmeshbbw}{{\bf {\it TBD \ }}}
\newcommand{\meshlatency}{{\bf {\it TBD \ }}}

\newcommand{\ecmbw}{0.96 TB/s}
\newcommand{\ocmbw}{10.24 TB/s}
\newcommand{\ecmlatency}{20 ns}
\newcommand{\ocmlatency}{20 ns}

In the simulation, our chief goal is to understand the performance implications of the on-stack
network and the off-stack memory design.
Our simulator has three network configuration options:
\begin{itemize}
\item{XBar} -- An optical crossbar (as described in Section~\ref{sec_onstack_interconnect}),
with bisection bandwidth of \xbarbbw,
and maximum signal propagation time of \xbarlatency~clocks.
\item{HMesh} --
An electrical 2D mesh with bisection bandwidth 1.28 TB/s and
per hop signal latency (including forwarding and signal propagation time) of 5 clocks.
\item{LMesh} --
An electrical 2D mesh with bisection bandwidth 0.64 TB/s and
per hop signal latency (including forwarding and signal propagation time) of 5 clocks.
\end{itemize}
The two meshes employ dimension-order wormhole routing~\cite{deadlock-free}.
We estimated a worst-case power of 26 W for the optical crossbar.  Since
many components of the optical system power are fixed (e.g., laser,
ring trimming, etc.), we conservatively assumed a continuous power
of 26 W for the XBar.  We assumed an electrical energy of 196 pJ per transaction
per hop, including router overhead.  This aggressively assumes low swing
busses and ignores all leakage power in the electrical meshes.

We also simulate two memory interconnects, the OCM
interconnect (as described in Section \ref{sec_optmem})
plus an electrical interconnect:
\begin{itemize}
\item{OCM} --
Optically connected memory;
off-stack memory bandwidth is
\ocmbw, memory latency is \ocmlatency.
\item{ECM} --
Electrically connected memory;
off-stack memory bandwidth is
\ecmbw, memory latency is \ecmlatency.
\end{itemize}
The electrical memory interconnect is based on the ITRS roadmap,
according to which it will be impossible to implement an ECM
with performance equivalent to the proposed OCM.
Table~\ref{tab_extmem} contrasts the memory interconnects.

\begin{table}[t!]
\small
\center{
\begin{tabular}{l|c|c}
{ Resource} & OCM & ECM \\\hline
Memory controllers    & 64 & 64  \\
External connectivity & 256 fibers & 1536 pins \\
Channel width	      & 128 b half duplex & 12 b full duplex \\
Channel data rate     & 10 Gb/s & 10 Gb/s \\
Memory bandwidth      & 10.24 TB/s & 0.96 TB/s \\
Memory latency	      & 20 ns & 20 ns \\
\end{tabular}
}
\caption{Optical vs Electrical Memory Interconnects}
\label{tab_extmem}
\end{table}

We simulate five combinations: XBar/OCM (i.e. Corona), HMesh/OCM, LMesh/OCM,
HMesh/ECM, and LMesh/ECM.  These choices highlight, for each benchmark,
the performance gain, if any, due to faster memory and due to faster interconnect.
We ran each simulation for a
predetermined number of network requests (L2 misses).
These miss counts are shown in Table~\ref{tab_bmarks}.

%% file: eval.tex
\section{Performance Evaluation}
\label{sec_eval}

\begin{figure}[ht!]
\center
\includegraphics[width=.475\textwidth]{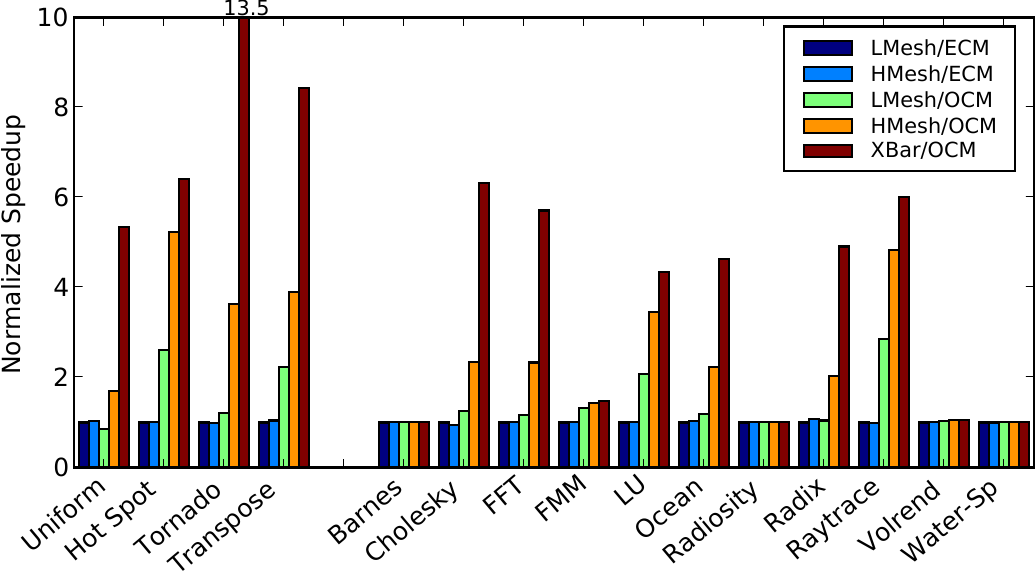}
\caption{Normalized Speedup}
\label{fig_speedup}
\end{figure}

For the five system configurations, Figure ~\ref{fig_speedup}
shows performance relative to the
realistic, electrically connected LMesh/ECM system.

We can form a few hypotheses based on the synthetic benchmarks.
Evidently, with low memory bandwidth, the high-performance mesh adds little value.
With fast OCM, there is very substantial performance gain over ECM systems, when using
the fast mesh or the crossbar interconnect, but much less gain if the low performance mesh is
used.  Most of the performance gain made possible by OCM is realized
only if the crossbar interconnect is used.
In the exceptional case, Hot Spot,
memory bandwidth remains the performance limiter
(because all the memory traffic is channeled
through a single cluster);
hence there is less pressure on the interconnect.
Overall, by moving to an OCM from an ECM in systems with an HMesh, we
achieve a geometric mean speedup of 3.28.  Adding the photonic
crossbar can provide a further speedup of 2.36 on the
synthetic benchmarks.

For the SPLASH-2 applications, we find that in four cases (Barnes,
Radiosity, Volrend, and Water-Sp) the LMesh/ECM system is
fully adequate.  These applications perform well due to their
low cache-miss rates and consequently low main memory bandwidth demands.
FMM is quite similar to these.  
The remaining applications
are memory bandwidth limited on ECM-based systems.
For Cholesky, FFT, Ocean, and Radix, fast memory provides considerable 
benefits, which are realized only with the fast crossbar.
LU and Raytrace are like Hot Spot: while OCM gives most of the 
significant speedup, some additional benefit derives from the 
use of the fast crossbar.
We posit below a possible reason for the difference
between Cholesky, FFT, Ocean, and Radix on the one hand, and LU and Raytrace on the other,
when examining the bandwidth and latency data.
These observations are generally consistent with the detailed memory traffic
measurements reported by Woo {\em et al.}~\cite{splash2}.
Overall, replacing an ECM with an OCM in a system using an
HMesh can provide a geometric mean speedup of 1.80.
Adding the photonic crossbar can provide a further speedup
of 1.44 on the SPLASH-2 applications.

\begin{figure}[ht!]
\center
\includegraphics[width=.475\textwidth]{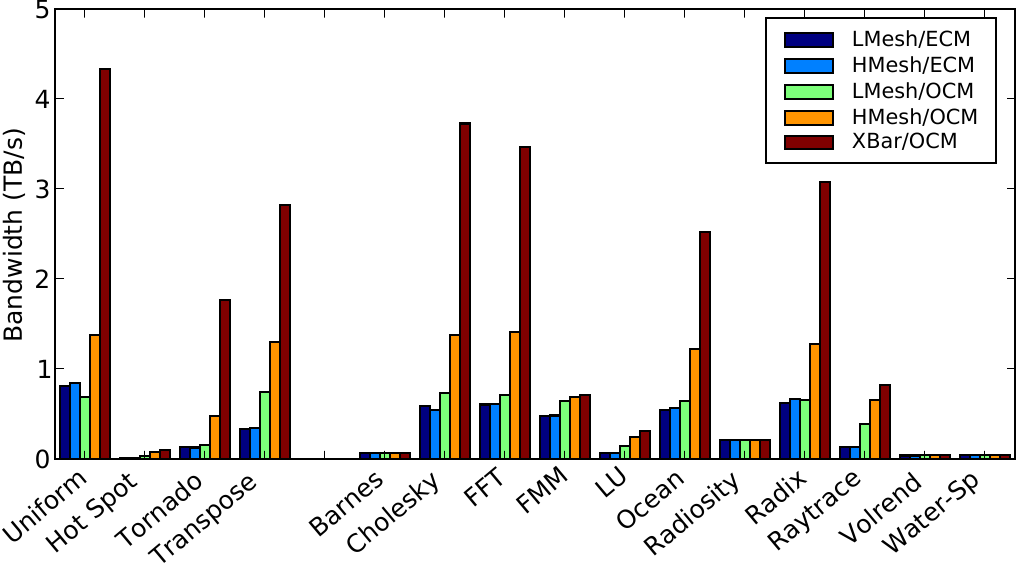}
\caption{Achieved Bandwidth}
\label{fig_bandwidth}
\end{figure}

Figure~\ref{fig_bandwidth} shows the actual rate of of communication with main memory.
The four low bandwidth applications that perform well on the LMesh/ECM configuration
are those with bandwidth demands lower than that provided by ECM.
FMM needs somewhat more memory bandwidth than ECM provides.
Three of the synthetic tests and four of the applications have
very high bandwidth and interconnect requirements,
in the 2 -- 5 TB/s range; these benefit the most from the XBar/OCM configuration.
LU and Raytrace do much better on OCM systems than ECM,
but do not require much more bandwidth than ECM provides. They appear to benefit
mainly from the improved latency offered by XBar/OCM.
We believe that this is due to bursty memory
traffic in these two applications.
Analysis of the LU code shows that many
threads attempt to access the same remotely stored matrix block
at the same time, following a barrier.
In a mesh, this oversubscribes the links into the
cluster that stores the requested block.

\begin{figure}[ht!]
\center
\includegraphics[width=.475\textwidth]{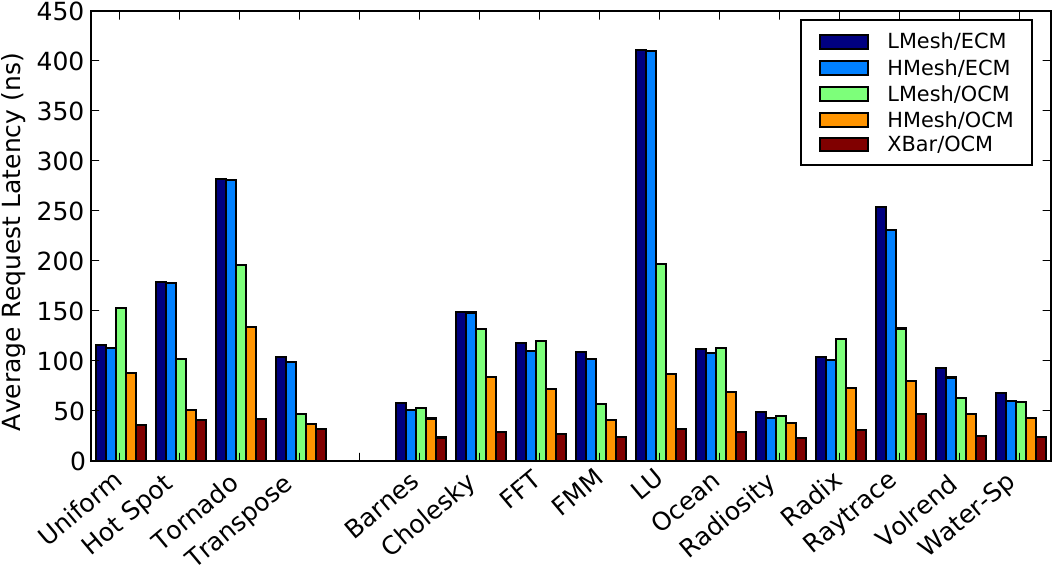}
\caption{Average L2 Miss Latency}
\label{fig_latency}
\end{figure}

Figure~\ref{fig_latency} reports the average latency of an L2 cache miss to main memory.
An L2 miss may be delayed in waiting for crossbar
arbitration (the token) and by flow-control (destination cluster buffers may be full) before
an interconnect message is generated.
Our latency statistics measure both queue waiting times and interconnect transit times.
LU and Raytrace see considerable average latency in ECM systems; it is improved dramatically
by OCM and improved further by the optical crossbar.
Note that the average latency can
be high even when overall bandwidth is low when traffic is bursty.

\begin{figure}[ht!]
\center
\includegraphics[width=.475\textwidth]{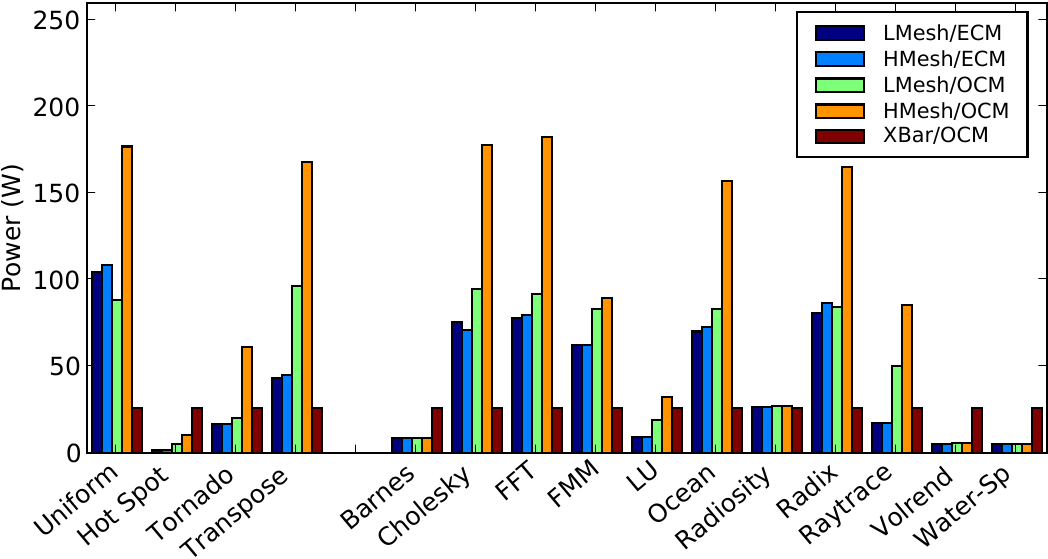}
\caption{On-chip Network Power}
\label{fig_power}
\end{figure}

Figure~\ref{fig_power} shows the on-chip network dynamic power.  For applications
that fit in the L2 cache, the photonic crossbar can dissipate more power
than for the electronic meshes (albeit ignoring mesh leakage power).
However, for applications with significant memory demands, the power
of the electronic meshes can fast become prohibitive with power of
100 W or more, even while providing lower performance.

%% file: related.tex
\section{Related Work}
\label{sec_related}


Recent proposals for optical networks in chip multiprocessors include
a hierarchical multi-bus with optical global layer~\cite{cornell}
and an optical, circuit-switched, mesh-based network
managed by electrical control packets~\cite{keren_hoti}.
In contrast, our crossbar interconnect uses optical arbitration and control.

Optical crossbars have been proposed for Asynchronous Transfer Mode (ATM)
switches~\cite{atm_xbar}
and for communication within processor clusters in large distributed shared
memory systems~\cite{dsm_xbar}.
This work relies on expensive VCSELs and free-space
optical gratings to demultiplex the crossbar's wavelengths, unlike
our solution which can be integrated into a modern 3D chip stack.

Most recent optical CMP interconnect proposals rely on electrical arbitration
techniques~\cite{petabit, cornell, keren_hoti}.
Optical arbitration techniques have been investigated in
SMP and ATM designs~\cite{p2p_token, symnet}.
While these techniques employ token rings, their tokens
circulate more slowly, as they are designed to
stop at every node in the ring,
whether or not the node is participating in the arbitration.

Chang et al.~\cite{rfi} overlay a 2D mesh CMP interconnect with a radio
frequency interconnect to provide low latency shortcuts.  They suggest
frequency division multiple access similar to our DWDM to provide multiple
channels per waveguide.
Capacitive~\cite{proximity} and inductive~\cite{inductive} coupling
technologies can provide wireless chip-to-chip communication which can
be used within a package.

The 8- and 16-core Sun Niagara~\cite{Niagara} and
Niagara2~\cite{niagara2} chips
use electrical crossbars.
The 80-core Intel Polaris chip~\cite{intel_polaris}
and the 64-core MIT Raw processor~\cite{raw}
connect their 
cores with 2D mesh networks.
A 2D mesh is easily laid out, regardless of its size.
Latency for nonlocal traffic is high because multiple hops are
required to communicate between cores unless they are physically
adjacent.   Random traffic is choked by the limited bisection bandwidth of
a mesh ($O(n)$ in an $n^2$-node mesh).
Express Virtual Channels (EVCs)~\cite{evc} alleviate the per-hop latency of
packet based mesh and torus networks,
but the paths cannot be arbitrarily shaped.

%% file: conclusion.tex
\section{Conclusions}
\label{sec_conclusion}

Over the coming decade, memory and
inter-core bandwidths must scale by orders of magnitude to support the
expected growth in per-socket core performance resulting from increased
transistor counts and device performance.  We believe recent developments in
nanophotonics can be crucial in providing required bandwidths
at acceptable power levels.

To investigate the potential benefits
of nanophotonics on computer systems we have developed an architectural
design called Corona.  Corona uses optically connected memories (OCMs)
that have been architected for low power and high bandwidth.  A set of
64 OCMs can provide 10 TB/s of memory bandwidth through 128 fibers using
dense wavelength division multiplexing.  Once this memory bandwidth comes
on chip, the next challenge is getting each byte to the right core out of
the hundreds on chip.  Corona uses a photonic crossbar with optical arbitration
to fully interconnect its cores, providing near uniform latency and
20 TB/s of on-stack bandwidth.

We simulated a 1024 thread Corona system running synthetic benchmarks
and scaled versions of the SPLASH-2 benchmark suite.
We found systems using
optically-connected memories and an optical crossbar between cores
could perform 2 to 6 times better on memory-intensive workloads
than systems using only electrical interconnects,
while dissipating much less interconnect power.
Thus we believe nanophotonics can be a compelling solution to both the
memory and network-on-chip bandwidth walls, while
simultaneously ameliorating the power wall.


%% file: npa.bbl
\begin{thebibliography}{10}
\providecommand{\url}[1]{#1}
\csname url@samestyle\endcsname
\providecommand{\newblock}{\relax}
\providecommand{\bibinfo}[2]{#2}
\providecommand{\BIBentrySTDinterwordspacing}{\spaceskip=0pt\relax}
\providecommand{\BIBentryALTinterwordstretchfactor}{4}
\providecommand{\BIBentryALTinterwordspacing}{\spaceskip=\fontdimen2\font plus
\BIBentryALTinterwordstretchfactor\fontdimen3\font minus
  \fontdimen4\font\relax}
\providecommand{\BIBforeignlanguage}[2]{{%
\expandafter\ifx\csname l@#1\endcsname\relax
\typeout{** WARNING: IEEEtranS.bst: No hyphenation pattern has been}%
\typeout{** loaded for the language `#1'. Using the pattern for}%
\typeout{** the default language instead.}%
\else
\language=\csname l@#1\endcsname
\fi
#2}}
\providecommand{\BIBdecl}{\relax}
\BIBdecl

\bibitem{3D_wkshop}
``{Proceedings of the ISSCC Workshop F2: Design of 3D-Chipstacks},'' IEEE, Feb
  2007, organizers: W. Weber and W. Bowhill.

\bibitem{802p5}
ANSI/IEEE, ``{Local Area Networks: {Token} Ring Access Method and Physical
  Layer Specifications, Std 802.5},'' Tech. Rep., 1989.

\bibitem{ucb_landscape06}
\BIBentryALTinterwordspacing
K.~Asanovic, R.~Bodik, B.~C. Catanzaro, J.~J. Gebis, P.~Husbands, K.~Keutzer,
  D.~A. Patterson, W.~L. Plishker, J.~Shalf, S.~W. Williams, and K.~A. Yelick,
  ``{The Landscape of Parallel Computing Research: A View from Berkeley},''
  EECS Department, University of California, Berkeley, Tech. Rep.
  UCB/EECS-2006-183, Dec 2006. [Online]. Available:
  \url{http://view.eecs.berkeley.edu}
\BIBentrySTDinterwordspacing

\bibitem{m5}
N.~L. Binkert, R.~G. Dreslinski, L.~R. Hsu, K.~T. Lim, A.~G. Saidi, and S.~K.
  Reinhardt, ``{The {M5} Simulator: Modeling Networked Systems},'' \emph{IEEE
  Micro}, vol.~26, no.~4, Jul/Aug 2006.

\bibitem{rfi}
\BIBentryALTinterwordspacing
M.-C.~F. Chang, J.~Cong, A.~Kaplan, M.~Naik, G.~Reinman, E.~Socher, and S.-W.
  Tam, ``{CMP Network-on-Chip Overlaid With Multi-Band RF-Interconnect},'' in
  \emph{HPCA}, Feb 2008. [Online]. Available:
  \url{http://www.gigascale.org/pubs/1225.html}
\BIBentrySTDinterwordspacing

\bibitem{petabit}
H.~J. Chao, K.-L. Deng, and Z.~Jing, ``{A Petabit Photonic Packet Switch
  (P3S)},'' in \emph{INFOCOM}, 2003.

\bibitem{chaudhry05}
S.~Chaudhry, P.~Caprioli, S.~Yip, and M.~Tremblay, ``{High-Performance
  Throughput Computing},'' \emph{IEEE Micro}, vol.~25, no.~3, May/Jun 2005.

\bibitem{p2p_token}
J.~Choi, H.~Lee, H.~Hong, H.~Kim, K.~Kim, and H.~Kim, ``Distributed optical
  contention resolution using an optical token-ring,'' \emph{Photonics
  Technology Letters, IEEE}, vol.~10, no.~10, Oct 1998.

\bibitem{deadlock-free}
W.~Dally and C.~Seitz, ``{Deadlock-Free Message Routing in Multiprocessor
  Interconnection Networks},'' \emph{IEEE Transactions on Computers}, vol.
  C-36, no.~5, May 1987.

\bibitem{proximity}
R.~Drost, R.~Hopkins, R.~Ho, and I.~Sutherland, ``{Proximity Communication},''
  \emph{JSSC}, vol.~39, no.~9, Sep 2004.

\bibitem{falconISPASS07}
A.~Falcon, P.~Faraboschi, and D.~Ortega, ``{Combining Simulation and
  Virtualization through Dynamic Sampling},'' in \emph{{ISPASS}}, Apr 2007.

\bibitem{silverthorne}
Intel, ``{Intel Atom Processor},'' http://www.intel.com/techno-logy/atom.

\bibitem{penryn}
------, ``{Introducing the 45nm Next Generation Intel Core
  Microarchitecture},'' http://www.intel.com/technology/magazine/
  45nm/coremicroarchitecture-0507.htm.

\bibitem{virtual_multicast}
N.~E. Jerger, L.-S. Peh, and M.~H. Lipasti, ``{Virtual Circuit Tree
  Multicasting: A Case for On-Chip Hardware Multicast Support},'' \emph{ISCA},
  Jun 2008.

\bibitem{cornell}
N.~Kirman, M.~Kirman, R.~K. Dokania, J.~F. Martinez, A.~B. Apsel, M.~A.
  Watkins, and D.~H. Albonesi, ``{Leveraging Optical Technology in Future
  Bus-based Chip Multiprocessors},'' in \emph{MICRO}, 2006.

\bibitem{bowers}
B.~R. Koch, A.~W. Fang, O.~Cohen, and J.~E. Bowers, ``{Mode-locked silicon
  evanescent lasers},'' \emph{Optics Express}, vol.~15, no.~18, Sep 2007.

\bibitem{Niagara}
P.~Kongetira, K.~Aingaran, and K.~Olukotun, ``{Niagara: A 32-Way Multithreaded
  Sparc Processor},'' \emph{IEEE Micro}, vol.~25, no.~2, 2005.

\bibitem{evc}
A.~Kumar, L.-S. Peh, P.~Kundu, and N.~K. Jha, ``{Express Virtual Channels:
  Towards the Ideal Interconnection Fabric},'' in \emph{{ISCA}}, Jun 2007.

\bibitem{kumar_tullsen}
R.~Kumar, V.~Zyuban, and D.~M. Tullsen, ``{Interconnections in Multi-Core
  Architectures: Understanding Mechanisms, Overheads and Scaling},'' in
  \emph{ISCA}, Jun 2005.

\bibitem{lipson05}
M.~Lipson, ``{Guiding, Modulating, and Emitting Light on Silicon--Challenges
  and Opportunities},'' \emph{Journal of Lightwave Technology}, vol.~23,
  no.~12, Dec 2005.

\bibitem{symnet}
\BIBentryALTinterwordspacing
A.~Louri and A.~K. Kodi, ``{SYMNET: An Optical Interconnection Network for
  Scalable High-Performance Symmetric Multiprocessors},'' \emph{Applied
  Optics}, vol.~42, no.~17, Jun 2003. [Online]. Available:
  \url{http://ao.osa.org/abstract.cfm?URI=ao-42-17-3407}
\BIBentrySTDinterwordspacing

\bibitem{adaptive_snooping}
M.~M.~K. Martin, D.~J. Sorin, M.~D. Hill, and D.~A. Wood, ``{Bandwidth Adaptive
  Snooping},'' in \emph{{HPCA}}, Feb 2002.

\bibitem{inductive}
N.~Miura, D.~Mizoguchi, T.~Sakurai, and T.~Kuroda, ``{Analysis and Design of
  Inductive Coupling and Transceiver Circuit for Inductive Inter-Chip Wireless
  Superconnect},'' \emph{JSSC}, vol.~40, no.~4, Apr 2005.

\bibitem{niagara2}
U.~Nawathe, M.~Hassan, L.~Warriner, K.~Yen, B.~Upputuri, D.~Greenhill,
  A.~Kumar, and H.~Park, ``{An 8-Core 64-Thread 64b Power-Efficient SPARC
  SoC},'' in \emph{ISSCC}, Feb 2007.

\bibitem{rambus2007tranceiver}
R.~Palmer, J.~Poulton, W.~J. Dally, J.~Eyles, A.~M. Fuller, T.~Greer,
  M.~Horowitz, M.~Kellam, F.~Quan, and F.~Zarkeshvarl, ``{A 14mW 6.25Gb/s
  Transceiver in 90nm CMOS for Serial Chip-to-Chip Communications},'' in
  \emph{ISSCC}, Feb 2007.

\bibitem{terabus}
L.~Schares, J.~Kash, F.~Doany, C.~Schow, C.~Schuster, D.~Kuchta,
  P.~Pepeljugoski, J.~Trewhella, C.~Baks, R.~John, L.~Shan, Y.~Kwark, R.~Budd,
  P.~Chiniwalla, F.~Libsch, J.~Rosner, C.~Tsang, C.~Patel, J.~Schaub,
  R.~Dangel, F.~Horst, B.~Offrein, D.~Kucharski, D.~Guckenberger, S.~Hegde,
  H.~Nyikal, C.-K. Lin, A.~Tandon, G.~Trott, M.~Nystrom, D.~Bour, M.~Tan, and
  D.~Dolfi, ``{Terabus: Terabit/Second-Class Card-Level Optical Interconnect
  Technologies},'' \emph{IEEE Journal of Selected Topics in Quantum
  Electronics}, vol.~12, no.~5, Sep/Oct 2006.

\bibitem{ITRS}
{Semiconductor Industries Association}, ``{International Technology Roadmap for
  Semiconductors. http://www.itrs.net/},'' 2006 Update.

\bibitem{keren_hoti}
A.~Shacham, B.~G. Lee, A.~Biberman, K.~Bergman, and L.~P. Carloni, ``{Photonic
  NoC for DMA Communications in Chip Multiprocessors},'' in \emph{IEEE Hot
  Interconnects}, Aug 2007.

\bibitem{raw}
M.~B. Taylor, W.~Lee, J.~Miller, D.~Wentzlaff, I.~Bratt, B.~Greenwald,
  H.~Hoffmann, P.~Johnson, J.~Kim, J.~Psota, A.~Saraf, N.~Shnidman,
  V.~Strumpen, M.~Frank, S.~Amarasinghe, and A.~Agarwal, ``{Evaluation of the
  Raw Microprocessor: An Exposed-Wire-Delay Architecture for ILP and
  Streams},'' in \emph{ISCA}, Jun 2004.

\bibitem{cacti5}
S.~Thoziyoor, N.~Muralimanohar, J.~Ahn, and N.~P. Jouppi, ``{CACTI} 5.1,'' HP
  Labs, Tech. Rep. HPL-2008-20.

\bibitem{intel_polaris}
S.~Vangal, J.~Howard, G.~Ruhl, S.~Dighe, H.~Wilson, J.~Tschanz, D.~Finan,
  P.~Iyer, A.~Singh, T.~Jacob, S.~Jain, S.~Venkataraman, Y.~Hoskote, and
  N.~Borkar, ``{An 80-Tile 1.28TFLOPS Network-on-Chip in 65nm CMOS},'' in
  \emph{ISSCC}, Feb 2007.

\bibitem{dsm_xbar}
\BIBentryALTinterwordspacing
B.~Webb and A.~Louri, ``{A Class of Highly Scalable Optical Crossbar-Connected
  Interconnection Networks (SOCNs) for Parallel Computing Systems},''
  \emph{TPDS}, vol.~11, no.~5, 2000. [Online]. Available:
  \url{citeseer.ist.psu.edu/webb00class.html}
\BIBentrySTDinterwordspacing

\bibitem{atm_xbar}
------, ``{All-Optical Crossbar Switch Using Wavelength Division Multiplexing
  and Vertical-Cavity Surface-Emitting Lasers },'' \emph{Applied Optics},
  vol.~38, no.~29, Oct 1999.

\bibitem{splash2}
S.~C. Woo, M.~Ohara, E.~Torrie, J.~P. Singh, and A.~Gupta, ``{The {SPLASH-2}
  Programs: Characterization and Methodological Considerations},'' in
  \emph{ISCA}, Jun 1995.

\bibitem{qianfan}
Q.~Xu, B.~Schmidt, S.~Pradhan, and M.~Lipson, ``{Micrometre-scale silicon
  electro-optic modulator},'' \emph{Nature}, vol. 435, May 2005.

\end{thebibliography}
